\documentclass[prl,twocolumn,aps,amsmath,reprint,nofootinbib,bibnotes,amsfonts,amssymb,superscriptaddress]{revtex4-1}
\usepackage[utf8]{inputenc}
\usepackage{amsmath,amssymb,amsfonts}
\usepackage{MnSymbol}
\def\Rbol{{\stackrel{\circ}{R}}{}}
\def\Rbolcal{{\stackrel{\circ}{\mathcal{R}}}{}}
\def\Sag{{\mathcal S}{}}
\def\Kag{{\mathcal K}{}}
\def\Mag{{\mathcal M}{}}
\def\Nag{{\mathcal N}{}}
\def\ombol{{\stackrel{\circ}{\omega}}{}}
\def\tref{{}^{0}h}

\begin{document}
\title{\bf Self-Excited Gravitational Instantons}
\author{Martin Kr\v{s}\v{s}\'ak}
\email{\texttt{martin.krssak@gmail.com, martin.krssak@fmph.uniba.sk}}
\affiliation{Department of Theoretical Physics, Faculty of Mathematics, Physics and Informatics, Comenius University in Bratislava, 84248, Slovak Republic}
\affiliation{Department of Astronomy, School of Physical Sciences, University of Science and Technology of China, 96 Jinzhai Road, Hefei, Anhui 230026, China}
\date{\today}
\begin{abstract}
We present a novel approach to constructing gravitational instantons based on the observation that the gravitational action of general relativity in its teleparallel formulation can be expressed as a product of the torsion and excitation forms. We introduce a new class of solutions where these two forms are equal, which we term the self-excited instantons, and advocate for their use over the self-dual instantons of Eguchi and Hanson. These new self-excited instantons exhibit striking similarities to BPST instantons in Yang-Mills theory, as their action reduces to a topological Nieh-Yan term, which allows us to identify the axial torsion as a topological current and show that the gravitational action is given by a topological charge.
\end{abstract}
\date{ \today}
\maketitle

\section{Introduction}
The finite Euclidean action solutions, known as instantons,  play an important role in  understanding the non-perturbative aspects of quantum field theories
\cite{Belavin:1975fg,Coleman:1978ae,Vainshtein:1981wh}. In the Yang-Mills case, the well-known BPST construction is based on considering gauge fields with  self-dual field strength, which automatically solve the field equations. The importance of these self dual instantons  lies in the fact that their action reduces to a topological term, revealing their topological nature, which led to a number of important insights into non-trivial vacuum structure of gauge theories and established the use of topology in the study of gauge field theories \cite{Callan:1976je,tHooft:1976snw,Eguchi:1980jx,Nash:1983cq,Manton:2004tk}.

Gravitational instantons were first introduced in the path integral derivation of black hole entropy \cite{Gibbons:1976ue}, followed by the proposal of self-dual gravitational instantons by Hawking  \cite{Hawking:1976jb}, which were further developed by Eguchi and Hanson, who constructed a new explicit self-dual solution \cite{Eguchi:1978gw,Eguchi:1978xp,Eguchi:1979yx}. Since then, gravitational instantons have been systematically studied and have provided important insights into a range of topics, including the non-perturbative structure of quantum gravity, gravitational anomalies, and black hole thermodynamics \cite{Gibbons:1978tef,Gibbons:1979xm,Gibbons:1979xn,Plebanski:1975wn,Witten:1985xe,Kronheimer:1989zs,Dunajski:2010zz}. While they share some similarities with the BPST instantons, the main limitation of this analogy is the fact that the gravitational action is not associated with any topological term in the Riemannian geometry.

In this work, we aim to demonstrate that,  within the teleparallel formulation of general relativity \cite{Aldrovandi:2013wha,Krssak:2018ywd,Krssak:2024xeh}, it is possible to find a new type of \textit{self-excited} gravitational instantons.
Our construction is based on the observation that the teleparallel action can be expressed as a product of torsion and excitation forms, making it natural to consider solutions for which these forms are equal. We then  argue that this is not only a fundamentally novel way  to construct gravitational instantons, but also that these self-excited instantons  are a closer analogue of the BPST construction, as their action is given by the well-known Nieh-Yan term \cite{Nieh:1981ww}, a topological invariant absent in the standard Riemannian geometry.

\section{Self-dual Instantons in YM Theory and General Relativity}
Let us briefly review the construction of self-dual instantons in Yang-Mills theory and general relativity. 
In the Yang-Mills case \cite{Nash:1983cq,Manton:2004tk}, we consider a  gauge connection $A$ with a field strength $F=DA=dA + A\wedge A$,
which satisfies the field equations
\begin{equation}\label{ymeq}
D F=0, \qquad
D\star F=0,	
\end{equation}
where  $\star$ is the Hodge dual operator. The first equation is the Bianchi identity, which the field strength $F$ automatically satisfies, and the second one is derived from the action
\begin{equation}\label{ymaction}
\Sag_\text{YM}=\int_\Mag \text{Tr}\, F\wedge\star F.	
\end{equation}

The (anti) self-dual  instantons have an (anti) self-dual field strength $F=\pm \star F$  \cite{Belavin:1975fg},  which obeys the Bianchi identity and consequently satisfies the second field equation as well. This simplifies the process of solving the field equations \eqref{ymeq}, as instead of solving them directly, we can solve much simpler algebraic condition of (anti) self-duality.

While this simplification is useful, the truly intriguing aspect of the BPST solution is that the action \eqref{ymaction} reduces to a topological term
\begin{equation}\label{ymdualac}
	\tilde{\Sag}_\text{YM}=\pm\int_\Mag \,\text{Tr} \,F\wedge F=
	\pm\int_{\Mag} d \Kag=\pm 8 \pi^2 k,
\end{equation}
where $\Kag$ is the Chern-Simons form. Applying Stokes' theorem, we find that the integral of $\Kag$  over well-behaved connections at infinity is 
equal to the integer $k$, known as the winding or instanton number. 
Moreover, this  turns out to be an absolute minimum of the action \eqref{ymaction}
\begin{equation}\label{bps}
S_\text{YM}\geq 8 \pi^2 |k|,
\end{equation}
known as the BPS bound, and hence  self-dual instantons  represent the dominant contribution to the path integral.

To show the construction of self-dual instantons in general relativity, we first review the  Riemannian geometry in Cartan formalims \cite{Fecko:2006zy}. The basic variables are the (co-)tetrad 1-forms $h^a=h^a{}_\mu dx^\mu$,  related to the metric tensor through $g_{\mu\nu}=\delta_{ab}h^a_{\ \mu} h^b_{\ \nu}$, where  $\delta_{ab}=\text{diag}(1,1,1,1)$ in the Euclidean case\footnote{Since the Latin indices are raised/lowered by the Euclidean metric $\delta_{ab}$, we do not have to care about their position, and summation convention applies whenever some index is repeated twice.}.
 
The  connection form  is a priori an independent variable, but in the case of the Riemannian connection\footnote{All geometric quantities with respect to the Riemannian connection are denoted  by $\circ$ above them.}  $\ombol^a{}_b$, it is fully determined from the tetrad by conditions of metric compatibility, implying $\ombol_{ab}=-\ombol_{ba}$, and vanishing torsion
\begin{equation}\label{zerotor}
	0= d h^a + \ombol^a{}_b \wedge h^b.
\end{equation}
The curvature 2-form $\Rbolcal^a{}_b=\frac{1}{2}\Rbol^a{}_{b\mu\nu} dx^\mu\wedge dx^\nu$,	
where $\Rbol^\alpha{}_{\beta\mu\nu}=h_a{}^\alpha h^b{}_\beta\Rbol^a{}_{b\mu\nu}$ are the components of the Riemannian curvature tensor,  is fully determined by
\begin{eqnarray}\label{curv}
	\Rbolcal^a{}_b&=& d\ombol^a{}_b +\ombol^a{}_c\wedge \ombol^c{}_b.
\end{eqnarray}

In general relativity,  gravity is described using the Riemannian geometry of spacetime, the dynamics of which is derived from the Hilbert action
	\begin{equation}\label{haction}
	\Sag_\text{EH}=-\int_\mathcal{M} h\, \Rbol d^4 x=-\int_\mathcal{M} \Rbolcal_{ab} \wedge \star (h^a\wedge h^b),	
\end{equation}
where $h=\det h^a{}_\mu$, $\Rbol=\Rbol^{\alpha\beta}{}_{\alpha\beta}$ is the Ricci scalar, in units $16\pi G/c^4=1$.

Following the  Eguchi-Hanson construction  \cite{Eguchi:1978gw,Eguchi:1978xp,Eguchi:1979yx}, we first introduce the dual curvature $	\smallstar \Rbolcal_{ab}=\frac{1}{2}\epsilon_{abcd} \Rbolcal_{cd}$, which allows us to write the Einstein field equations as 
\begin{equation}\label{einfe}
\smallstar \Rbolcal^a{}_b \wedge h^b  =0.
\end{equation}  
The (anti) self-dual curvature  $\Rbolcal_{ab}=\pm	\smallstar \Rbolcal_{ab}$ then automatically satisfies the Einstein field equations \eqref{einfe}   on the account of the Bianchi identity $\Rbolcal^a{}_b \wedge h^b  =0$, and  implies (anti) self-duality of the connection
\begin{equation}\label{sdconn}
\ombol_{ab}=\pm \smallstar\ombol_{ab}.
\end{equation} 
Eguchi and Hanson then considered an ansatz
\begin{equation}\label{ehtet}
h^a=(f d r,r \sigma_x, r \sigma_y, r g \sigma_z ),	
\end{equation}
where $\sigma_i$ are the $SU(2)$ Cartan-Maurer forms on $S^3$
\begin{flalign}
&\sigma_x=
\frac{1}{2}(\sin\psi\, d\theta-\sin\theta \cos\psi\, d\phi),\\
&\sigma_y=
\frac{1}{2}(-\cos\psi\, d\theta-\sin\theta \sin\psi\,d\phi),\\
&\sigma_z=
\frac{1}{2}(d\psi+\cos\theta\,d\phi),
\end{flalign}
and found a solution
\begin{equation}\label{ehsol}
f^2=g^{-2}=1-\frac{a^4}{r^4}.
\end{equation}

The process of finding this solution is  indeed analogous to the  BPST method:  instead of solving the field equations  \eqref{einfe} directly,  we solve  the self-duality condition \eqref{sdconn}, and our solution is then  guaranteed to automatically solve  the field equations as well. 

However,  there are some significant differences including  the fact that  $\smallstar$ is not the usual Hodge dual  but rather the ``internal" or ``tangent" space dual \cite{Baez:1995sj,Kol:2023yxd}. 
The key difference is that, while the BPST action is a topological term \eqref{ymdualac}, the full gravitational action--including the Gibbons-Hawking boundary terms and appropriate counterterms \cite{Gibbons:1976ue}--is not associated with any topological term in the Riemannian geometry. While there are two other topological terms, the Euler class $\chi$ and Pontryagin class $P_1$ (see Table~\ref{table1}), which can be used to classify instantons \cite{Eguchi:1978gw, Eguchi:1980jx},  these are quadratic in curvature and thus cannot be related to the action. As a consequence, we cannot discuss any topological aspects of the gravitational action, which, in our view, significantly limits the extent and usefulness of the analogy with the BPST construction.

\section{Teleparallel Formulation of General Relativity}
Teleparallel gravity is a reformulation of general relativity \cite{Krssak:2018ywd,Krssak:2024xeh}, where instead of the Riemannian connection $\ombol^a{}_b$, we use the teleparallel connection $\omega^a{}_b$, defined in a complimentary way as  a metric connection with vanishing curvature 
\begin{equation}\label{key}
d\omega^a{}_b +\omega^a{}_c\wedge \omega^c{}_b=0,	
\end{equation}
which has generally non-vanishing teleparallel torsion
\begin{eqnarray}\label{teletor}
T^a= d h^a + \omega^a{}_b \wedge h^b, 
\end{eqnarray}
and satisfies the Bianchi identity 
\begin{equation}\label{telebia}
DT^a\equiv d T^a +\omega^a{}_b \wedge T^b=0.	
\end{equation}
We can then consider $d(h^a\wedge T^a)$ and by straightforwardly applying \eqref{teletor} and \eqref{telebia} we find
\begin{equation}\label{nyid}
d(h^a\wedge T_a)=
T^a\wedge T_a,
\end{equation}
which is just the  Nieh-Yan identity \cite{Nieh:1981ww} for the flat teleparallel connection.

The teleparallel connection  is related to the Riemannian connection by the Ricci theorem
\begin{equation}\label{ricci}
\omega^a{}_b=\ombol^a{}_b+ K^a{}_b,
\end{equation}
where $K^a{}_b$ is the contortion 1-form related to the torsion as $T^a
=K^a{}_b \wedge h^b$. Using  \eqref{ricci} we can rewrite the Ricci scalar in terms of teleparallel geometric quantities as \cite{Aldrovandi:2013wha}
\begin{equation}\label{teleid}
-\Rbol=T+\frac{2}{h} \partial_\mu (T^{\nu\mu}{}_\nu),	
\end{equation}
where $T$ is the teleparallel torsion scalar
\begin{equation}\label{tscalar}
T=
\frac{1}{4}	T_{\rho\mu\nu}T^{\rho\mu\nu}
+\frac{1}{2}T_{\rho\mu\nu}T^{\nu\mu\rho}
-T^\nu{}_{\mu\nu}T^{\rho\mu}{}_{\rho}.
\end{equation}

The idea of teleparallel gravity is to consider an action given by the torsion scalar \eqref{tscalar}, which is guaranteed to be dynamically equivalent to general relativity since it differs from the Ricci scalar only by a  total derivative \eqref{teleid}. This can be understood as removing the problematic total derivative term hidden in the Hilbert action and geometrically covariantizing the remaining bulk Einstein action \cite{Krssak:2024kva,Krssak:2024xeh,Krssak:2023nrw}.
  
The key observation is that since the torsion scalar is quadratic in torsion, the teleparallel action can be naturally written as a product of two forms \cite{Obukhov:2002tm,Lucas:2009nq}
\begin{equation}\label{teleac}
\Sag_\text{TG}=\int_\mathcal{M} h T d^4 x=\int_\mathcal{M} T^a\wedge H_a,	
\end{equation}
where $H^a=\frac{1}{2} H^a{}_{\rho\sigma} dx^\rho\wedge dx^\sigma$  is the so-called excitation form, the components of which are  given by
\begin{equation}\label{exci}
H^a{}_{\rho\sigma}=h \epsilon_{\rho\sigma\alpha\beta}\left(\frac{1}{4}T^{a\alpha\beta}+\frac{1}{2}T^{\alpha a \beta}-h^{a\beta}T^{\nu\alpha}{}_\nu\right),	
\end{equation}
where $\epsilon_{\rho\sigma\alpha\beta}$ is the Levi-Civita symbol with $\epsilon_{0123}=1$.

Varying \eqref{teleac} with respect to $h^a$, we obtain the vacuum field equations 
\begin{eqnarray}
DH^a+ E^a&=&0\label{teleeq2}.
\end{eqnarray}
where $E^a$ is the gravitational energy-momentum current form, and variation with respect to $\omega^a{}_b$ is trivial \cite{Krssak:2015lba,Golovnev:2017dox} These equations are fully equivalent to the Einstein field equations, but have a form strikingly similar to the Yang-Mills theory.

\section{Self-Excited Instantons in Teleparallel Gravity}
Let us start with an observation  that self-duality is not  actually required to show that the self-dual Yang-Mills action is topological \eqref{ymdualac}.  The whole construction would work more generally if we would replace $\star F$ by $H$ in \eqref{ymaction}, considered $F=\pm H$, and \eqref{ymdualac} would still follow from the Bianchi identity. This goes  in the spirit of the premetric or axiomatic electromagnetism, which suggests to treat electromagnetism more generally in terms of $F$ and $H$, and then view the Maxwell electromagnetism only as a special case of the constitutive relation  $H=\star F$ \cite{HehlObukhov,Hehl:2016glb}.   

Motivated by this observation and the recent works suggesting that teleparallel gravity can be treated in a similar way \cite{Itin:2016nxk,Hohmann:2017duq,Itin:2018dru}, we  consider a new class of solutions where the torsion form is (up to a sign) equal to the excitation form. Instead of (anti) self-dual solutions, we have then (anti) self-excited solutions 
\begin{equation}\label{second}
	T^a=\pm \,H^a.
\end{equation}
It is obvious that we again obtain automatic solutions of the field equations since torsion automatically obeys the Bianchi identity and we have  $DT^a=\pm DH^a=0$. This also implies vanishing gravitational energy-momentum current $E^a=0$ for our (anti) self-excited solutions.
 
The novelty compared to Eguchi-Hanson is that  for  (anti) self-excited solutions  \eqref{second} the action \eqref{teleac} reduces
to  the Nieh-Yan term  \eqref{nyid}, and we can write it as
\begin{equation}\label{seact}
\tilde{\Sag}_\text{TG}=\pm \int_\mathcal{M} T^a\wedge T_a=\pm \int_\mathcal{M} d(h^a\wedge T_a)=\pm \Nag.
\end{equation}
The total derivative term  is the so-called axial torsion, which can be written in components as
\begin{equation}\label{key}
 d(h^a\wedge T_a)=\frac{1}{2} \partial_\mu (\epsilon^{\mu\nu\rho\sigma} T_{\nu\rho\sigma})d^4\,x=
 3\, (\partial_\mu a^\mu)\, d^4x ,
\end{equation}
where $a^\mu=\frac{1}{6}\epsilon^{\mu\nu\rho\sigma} T_{\nu\rho\sigma}$ \cite{Aldrovandi:2013wha}. 

Applying the Stokes' theorem, we find that the action is given only by the asymptotic value of the axial torsion, and  we denote its value  as the Nieh-Yan charge $\Nag$.
The axial torsion then plays the role of the Chern-Simons current and the Nieh-Yan charge is analogous to the instanton number.

\section{Examples}
We can verify that the Eguchi-Hanson instanton \eqref{ehtet}-\eqref{ehsol} is a self-excited solution. Here, it is essential to address the role of the teleparallel connection $\omega^a{}_b$ in teleparallel gravity. While this connection is not dynamical, it is relevant for determining the value of the action \cite{Krssak:2015rqa,Krssak:2024kva}. To illustrate this, we consider the ansatz \eqref{ehtet} in the so-called the Weitzenböck gauge $\omega^a{}_b = 0$ \cite{Obukhov:2002tm}, and  confirm that it is indeed a self-excited solution \eqref{second}. However, the axial torsion behaves as $\propto \mathcal{O}(r^2)$, causing the action to diverge. 

The well-known solution is to consider a reference tetrad $\tref^a=(d r,r \sigma_x, r \sigma_y, r\sigma_z )$ and calculate the teleparallel connection as the Riemannian connection of the reference tetrad $\omega^a{}_b=\ombol^a{}_b (\tref^a)$ \cite{Krssak:2015rqa,Krssak:2018ywd,Krssak:2024kva}. 
We can then check that the resulting torsion remains self-excited, but the asymptotic behavior of the axial torsion  is changed to
$ \propto \mathcal{O}(r^{-6})$, which vanishes at infinity. We then find that the Nieh-Yan charge vanishes and hence recover the Eguchi-Hanson result that the full gravitational action for their instanton is zero \cite{Eguchi:1978gw}. This is
 consistent with our recent argument  that the teleparallel action \eqref{teleac} with a suitably chosen teleparallel connection is fully equivalent to the full general relativity action including all boundary and counterterms \cite{Krssak:2024kva}. See \cite{Krssak:2023nrw,Krssak:2024xeh} as well.

An example  with a non-trivial topological charge can be obtained using $\omega^a{}_b =0$ and an ansatz  
\begin{equation}\label{ohtet}
	h^a=(f d r,g\, \sigma_x, g\, \sigma_y, g\, \sigma_z ),
\end{equation}
which was considered in \cite{Chandia:1997hu}  with $f=r^{-1},g=1$,  while in \cite{Obukhov:1997pz} authors have used $2f=\pm g'$, obtained from the (anti) self-duality condition $T^a=\pm\star T^a$. While both these choices of $f$ and $g$ do lead to a non-trivial Nieh-Yan charge \eqref{seact}, the problem is that they are not actually solutions of the field equations.

In contrast, our (anti) self-excitement  condition  \eqref{second} coincidentally implies  a similar result  $f=\pm g'$,  which, crucially, does solve the field equations \eqref{teleeq2}. We  find that the axial torsion  can be written as
\begin{equation}\label{ohax}
h^a\wedge T_a= 6 g^2\, \sigma_x\wedge\sigma_y\wedge\sigma_z. 
\end{equation}
For all functions that goes as $g\rightarrow 1$ at $r\rightarrow\infty$, and using that the volume of $S^3$ is $\int_{S^3}\sigma_x\wedge\sigma_y\wedge\sigma_z=2\pi^2$, we obtain  the Nieh-Yan charge as
\begin{equation}\label{key}
\Nag= 12\, \pi^2,
\end{equation}
which  is the same result as obtained in  \cite{Chandia:1997hu} since the axial torsion \eqref{ohax} is independent of $f$. 
Note that the Nieh-Yan charge does not depend on the  functions $f$ and $g$, as expected from a topological charge. 

\onecolumngrid
\begin{center}
	\begin{table}[t]
		\begin{tabular}{ |l|l|l|l| } 
			\hline
			& \textbf{Yang-Mills Theory} & \textbf{General Relativity}&\textbf{Teleparallel Gravity} \\ 
			\hline
			Basic variables & $A$& $h^a$ & $h^a,\omega^a{}_b$ \\ 
			\hline
			Field strength & $F=DA=dA+A\wedge A$ & $\Rbolcal^a{}_b=d\ombol^a{}_b+\ombol^a{}_c\wedge\ombol^c{}_b$
			&$T^a=Dh^a=dh^a+\omega^a{}_b\wedge h^b$
			\\ 
			\hline
			Action & $\int \text{Tr} F\wedge \star F$ & $
			-\int \Rbolcal_{ab} \wedge \star (h^a\wedge h^b)	$ & $\int T^a \wedge H_a$
			\\ 
			\hline
			Bianchi identity & $DF=0$& $\Rbolcal^a{}_b\wedge h^b=0$ & $DT^a=0$ \\ 
			\hline
			Field equations & $D \star F=0$& $\smallstar\Rbolcal^a{}_b\wedge h^b=0$ 
			& $DH^a+E^a=0$ \\ 
			\hline
			Self-dual field strength  & $F=\pm \star F$ & $
			\Rbolcal_{ab}=\pm \smallstar \Rbolcal_{ab}$ & $T^a=\pm H^a$\\ 
			\hline
			Self-dual solution & $F=\pm \star F$ & $\ombol^a{}_b=\pm \smallstar \ombol^a{}_b$ & $T^a=\pm H^a$ \\ 
			\hline
			Topological term(s) & $\int \text{Tr} F\wedge F$ & $\int \epsilon_{abcd}\, \Rbolcal^{ab}\wedge \Rbolcal^{cd}$ & $\int T^a \wedge T_a$ \\
			
			&  & $\int \Rbolcal^{a}{}_b\wedge \Rbolcal^{b}{}_a$ & \\
			\hline 
			Topological charge(s) & $k$ & $\chi,P_1$ & $\Nag$ \\
			\hline 
		\end{tabular}
		\caption{A comparison of self-dual and self-excited solutions in Yang-Mills theory, general relativity, and teleparallel gravity. \label{table1}}
	\end{table}
\end{center}
\begin{center}
\rule{5cm}{0.4pt}
\end{center}
\twocolumngrid 

\section{Discussion and Conclusions}
Let us explain the relation between our results and previous works. The Nieh-Yan term was originally discovered in Riemann-Cartan geometry \cite{Nieh:1981ww}, which is the underlying geometry of Einstein-Cartan and Poincare gauge theories of gravity \cite{Hehl:1976kj, Hehl:1994ue, Blagojevic:2000pi,Obukhov:2002tm}. In these theories, torsion appears in addition to curvature, representing additional degrees of freedom compared to general relativity. The Nieh-Yan term then emerges as an additional topological term alongside the Euler and Pontryagin terms, and it is relevant for the axial anomaly because the axial torsion couples directly to fermionic fields in metric-affine gravity \cite{Obukhov:1982da}. 
Most of the previous works discussing torsional instantons or the Nieh-Yan term were conducted either in this context \cite{Obukhov:1997pz, Chandia:1997hu, Li:1999ue}, or various axion-like modified gravity theories \cite{Mielke:2006zp,Mielke:2009zz,Chatzistavrakidis:2020wum,Li:2020xjt,Hohmann:2020dgy,Chatzistavrakidis:2021oyp,Li:2021wij,Li:2023fto}. 

This differs significantly from our approach and overall philosophy, as we consider the teleparallel formulation of general relativity, where torsion represents the same degrees of freedom as in general relativity \cite{Krssak:2024xeh}. 
Nevertheless, these previous works are relevant to us, particularly \cite{Obukhov:1997pz, Chandia:1997hu, Li:1999ue}, where the authors have simplified their analysis of torsional instantons by setting curvature to zero, thereby reducing their Riemann-Cartan geometry to teleparallel geometry. 
However, with a primary motivation being the metric-affine gravity, their intention was to present just illustrative examples of the non-trivial Nieh-Yan term \eqref{nyid}, without specifying the action of the theory.
This explains why they calculated the Nieh-Yan term for some teleparallel torsion without addressing whether this torsion actually satisfies any field equations. 

In contrast, we discuss teleparallel gravity given by the action \eqref{teleac}, where torsion must satisfy the Einstein field equations in their teleparallel form \eqref{teleeq2}. In the case of \eqref{ohtet}, this results in the relation $f=\pm g'$, 
despite the fact that the resulting  the Nieh-Yan charge is independent of $f$ and $g$. 

Our work was originally motivated by the premetric approach to gravity \cite{Itin:2016nxk,Hohmann:2017duq,Itin:2018dru} and the claim of  new kind of self-duality introduced in discussion of linearized gravity \cite{deAndrade:2005xy}. This new self-duality was meant in terms of a novel ``soldered" duality operator $\oast$, which would allow  to write  the excitation form  as $H^a=\oast T^a$ \cite{Lucas:2008gs,Aldrovandi:2013wha}. However, the recent analysis has shown that this operator is not well-defined \cite{Hornak:2024dkj}, and our results here demonstrate that such an operator is actually not needed.

Indeed, our work suggests that the concept of self-excitement leads to a closer gravitational analogue of the BPST construction than  self-duality.  Our self-excited instantons not only automatically satisfy the field equations but their action is given by  a topological Nieh-Yan term. We anticipate that this approach will open up new perspective for studying the topological aspects of the gravitational action and, hopefully,  provide important insights similar to those obtained in Yang-Mills theory.

This leads us to the final question of the relevance of self-duality in instanton construction and whether it can be completely replaced by the concept of self-excited solutions. Returning to the Yang-Mills case, we can recognize  that self-duality  plays an important role in the  proof of the  BPS bound \eqref{bps}, i.e. we cannot replace $\star F$ by $H$ and  derive the BPS bound without additional assumptions.  

Therefore, we do not expect an analogue of the BPS bound for our (anti) self-excited solutions. While this somewhat limits their importance for the Euclidean quantization \cite{Hawking:1978jn}, as we cannot prove that they are the dominant contribution to the path integral, it is consistent with the well-known result that the gravitational action is not bounded from below \cite{Gibbons:1978ac}. We hope that our approach will provide a better understanding of this problem using the topological insights offered by our construction.

\section{Acknowledgements}
This work was funded through  SASPRO2 project \textit{AGE of Gravity: Alternative Geometries of Gravity}, which has received funding from the European Union's Horizon 2020 research and innovation programme under the Marie Skłodowska-Curie grant agreement No. 945478. 

\bibliography{BulkReferences}
\bibliographystyle{Style}
\end{document}